\documentclass{jpp}
\usepackage{graphicx}
\usepackage{epstopdf, epsfig}
\usepackage{hyperref}
\usepackage{amsmath,amsfonts,amssymb}
\usepackage{subfigure}

\usepackage{color}
\usepackage{cancel}
\usepackage{natbib}
\usepackage{upgreek}

\shorttitle{Machine-learning Closure for Vlasov-Poisson Dynamics}
\shortauthor{N. Barbour, W. Dorland, I. G. Abel, and M. Landreman}

\title{Machine-learning Closure for Vlasov-Poisson Dynamics in Fourier-Hermite Space}

\author{N. Barbour\aff{1,2}\corresp{\email{barbour@umd.edu}}, W. Dorland \aff{1,2}, I. G. Abel\aff{2}, and M. Landreman\aff{2}}
	
\affiliation{\aff{1}Department of Physics, University of Maryland, College Park, Maryland 20742, USA
\aff{2}Institute for Research in Electronics and Applied Physics, University of Maryland, College Park, Maryland 20742, USA
}

\newcommand{\pd}[2]{\frac{\partial #1}{\partial #2}}
\newcommand{\pdt}[1]{\frac{\partial #1}{\partial t}}
\newcommand{\pdv}[1]{\frac{\partial #1}{\partial v}}
\newcommand{\fft}[1]{\mathcal{F}[#1]}
\newcommand{\ifft}[1]{\mathcal{F}^{-1}[#1]}

\begin{document}

\maketitle

\begin{abstract}
	Accurate reduced models of turbulence are desirable to facilitate the optimization of magnetic-confinement fusion reactor designs.
	As a first step toward higher-dimensional turbulence applications, we use reservoir computing, a machine-learning (ML) architecture, to develop a closure model for a limiting case of electrostatic gyrokinetics.  
	We implement a pseudo-spectral Eulerian code to solve the one-dimensional Vlasov-Poisson system on a basis of Fourier modes in configuration space and Hermite polynomials in velocity space.  
	When cast onto the Hermite basis, the Vlasov equation becomes an infinitely coupled hierarchy of fluid moments, presenting a closure problem. 
	We exploit the locality of interactions in the Hermite representation to introduce an ML closure model of the small-scale dynamics in velocity space.  
	In the linear limit, when the kinetic Fourier-Hermite solver is augmented with the reservoir closure, the closure permits a reduction of the velocity resolution, with a relative error within two percent for the Hermite moment where the reservoir closes the hierarchy.	
	In the strongly-nonlinear regime, the ML closure model more accurately resolves the low-order Fourier and Hermite spectra when compared to a na\"{i}ve closure by truncation and reduces the required velocity resolution by a factor of sixteen.

\end{abstract}

\section{Introduction}
One of the most important ongoing efforts in the pursuit of magnetic fusion
energy is the study of microturbulence \citep{Yoshida25}.  
While magnetic-confinement fusion plasmas are designed to be stable to macroscopic magnetohydrodynamic (MHD) perturbations, 
the steep gradients in temperature and density that are present in all current designs 
also drive small-scale instabilities.  
These instabilities drive sustained turbulence, which then dominates the transport of particles, energy, and momentum in the plasma.
  
As the fusion industry turns its focus toward designing first-of-a-kind pilot plants, optimizing these designs to mitigate turbulent heat losses is an urgent objective.  
Significant progress in turbulence optimization for tokamaks \citep{Highcock18} and stellarators \citep{Roberg_Clark23, Kim24} has been achieved.
However, the space of candidate magnetic equilibria remains vast, and high-resolution turbulence simulations are computationally expensive.  
A unifying feature of fluid \citep{Kolmogorov41, Kraichnan59}, MHD \citep{Goldreich95}, and gyrokinetic \citep{BanonNavarro11} turbulence theory is that free energy tends to cascade from large scales to small scales through local interactions in phase space, and these cascades are well-described by power laws\footnote{ Power laws typically arise when a scale-local transfer of energy occurs, and a separation between energy injection and dissipation scales exists. The relevance of scale locality to energy cascades in gyrokinetics has been studied in detail by \citet{Barnes11} and \citet{Teaca17}. In some systems of interest, inverse cascades occur, driving invariants from small scales to large scales.  The theory of inverse cascades has been developed for several two-dimensional systems \citep{Hasegawa78,Plunk10}, yet an open question remains about their applicability to the full five-dimensional gyrokinetic system.}.
Therefore, models that can accurately calculate large-scale quantities, including spectra and heat fluxes, without needing to resolve small scales, are both desirable and plausible.

In the field of fluid turbulence modeling, two successful approaches to constructing reduced, lower-resolution models that capture large-scale behavior are Reynolds-Averaged Navier-Stokes (RANS, \citet{Reynolds1895}) and large eddy simulation (LES, \citet{Deardorff70}). 
LES also has been used to build reduced models of gyrokinetic turbulence \citep{BanonNavarro14}. 
These formalisms accelerate turbulence simulations by averaging over time and filtering out the small-scale dynamics. 
Closure models provide another pathway to filter out the small scales while retaining accurate large-scale statistics.  
In kinetic plasma physics, \citet{HammettPerkins} developed a landmark analytic linear closure for the low-frequency response to small Langmuir oscillations.

An alternative method for building reduced models is to use data-driven, statistical techniques like machine learning (ML) and Bayesian optimization.  
In recent years, the proliferation of ML models has led to advancements in pattern identification and model construction for magnetic fusion.  
Some example applications include improved prediction and avoidance of tokamak disruptions using the \texttt{DECAF} framework \citep{Piccione20} and on the DIII-D tokamak \citep{Rea19, Fu20}, improved active-feedback plasma control systems with reinforcement learning \citep{DeepMind22, Seo24, Kerboua-Benlarbi24}, optimized execution of gyrokinetic simulations with the \texttt{PORTALS} framework \citep{Rodriguez-Fernandez22}, and accelerated, low-resolution turbulence simulations \citep{Greif23, Castagna24, Clavier25}. 
Data-driven, configuration space \citep{Ma20, Qin23} and heat flux \citep{Ingelsten24, Huang25} closure models of Landau damping have also been developed. 

In this paper, we choose to use the one-dimensional, collisionless Vlasov-Poisson system as a test problem for implementing an ML closure in velocity space.  
This system is the one-dimensional, electrostatic limit of the full five-dimensional gyrokinetic system \citep{ParkerDellar}.  
It models dynamics parallel to the magnetic field, and it contains a quadratic nonlinearity present in the full gyrokinetic system.  
These features make the Vlasov-Poisson system a suitable foundation to assess the future potential for a closure model to be developed for gyrokinetic turbulence. 

In parallel velocity space, we use properties of the Hermite basis to cast the objective of reducing resolution requirements as a closure problem.  
The Hermite basis has been used extensively to solve problems in kinetic plasma dynamics \citep{Armstrong67, Grant67, Joyce71, Gibelli06, Zocco11, Loureiro13, ParkerDellar, Kanekar15, Jorge17, Adkins18, Mandell18, Frei23}.
In this work, we opt to focus on velocity space because the Hermite basis naturally expresses the interaction between scales in velocity space as local interactions in a coupled hierarchy of moments.  This property of the moment formalism directly presents a single-term target for a closure.   
Specifically, the interaction between resolved scales and unresolved scales in velocity space appears as a single unknown term in an equation in this representation.
In contrast, the spectral representation of configuration space introduces some nonlocal interactions across scales, complicating the development of a closure model.  
Additionally, the Hermite basis explicitly expresses conservation laws for particle number, momentum, and energy as evolution equations for the lowest-order moments. 
These properties allow us to develop a closure that preserves the low-order conservation laws.  More details on the closure problem are presented in Section \ref{sec:closure}\,.  

We use reservoir computing, an ML architecture, to implement a spectral, proof-of-concept, velocity-space closure for the moment hierarchy that maintains accurate spectra and reduces resolution requirements.  
The reservoir computing model maintains an internal representation of the system state, which it evolves through a directed graph with fixed weights.  
It then predicts the system state at time $t+1$ from the state at time $t$ using a set of output weights trained by linear regression.  
Since its concurrent development by \citet{Jaeger01} and \citet{Maass02}, reservoir computing has successfully been applied to the task of forecasting the dynamics of low-dimensional chaotic systems \citep{Lu17, PathakPRL}.
This ML paradigm is well-suited to the task of modeling time-series data of physical processes.
In climate physics, \citet{Arcomano20} built a surrogate global atmospheric forecast model with reservoir computing, and in plasma physics, \citet{Jalalvand22} used reservoir computing to analyze and classify Alfv\'{e}n eigenmodes using DIII-D diagnostics data.  
In comparison to several other recurrent neural network architectures, reservoir computing has shown comparable ability to capture the statistics of chaotic systems, while requiring significantly less training time \citep{Vlachas20}. 
These results motivate our decision to use reservoir computing for the closure model.

Some ML methods, including reservoir computing, are more challenging to interpret than traditional physics models. 
Artificial neural network architectures are often ``black boxes" that generate layered, analytically-intractable networks of nonlinear activation functions. Gaining insightful intuition from these networks can be daunting.  
We mitigate this problem by constraining the ML portion of our simulations to represent only the unresolved small scales, preserving analytic evolution equations for large scales. 
Our methods are analogous to an LES approach for phase space.  
  
The paper has the following structure.  In Section \ref{sec:VP}\,, we present our test problem, the normalized one-dimensional Vlasov-Poisson system.  We then derive the projection of the system onto the pseudo-spectral Fourier-Hermite Basis in Section \ref{sec:pseudo-spectral}\,.  We present the closure problem in Section \ref{sec:closure} and introduce our machine-learning closure model in Section \ref{sec:reservoir} as a solution to that problem.  In Section \ref{sec:results}\,, we present our ML model's predictions, and we conclude with a summary and discussion in Section \ref{sec:conclusion}\,.

\section{Vlasov-Poisson System}\label{sec:VP}
As our test problem, we study the dynamics of a one-dimensional, collisionless, electrostatic hydrogen plasma.  We focus on the electron dynamics, treating the ions as a cold, neutralizing background. The equations that describe this plasma are the Vlasov-Poisson system \citep{Vlasov}. The calculations in this paper use Gaussian units. The Vlasov equation for electrons in one dimension is 

\begin{equation}\label{eq:vlasov_unnormalized}
	\pdt{f} + v \pd{f}{z} -  \frac{eE}{m_e} \pdv{f} = 0,
\end{equation}
where $f(z,v,t)$ is the electron distribution function, $e$ is the elementary charge, and $m_e$ is the mass of an electron. $E(z,t)$ is the electric field, which we calculate using Poisson's equation,

\begin{equation}
	-\pd{^2\Phi}{z^2} = 4 \upi \rho, \quad E = -\pd{\Phi}{z},
\end{equation}
where $\rho(z,t)$ is the total electric charge density, including both ions and electrons, and $\Phi(z,t)$ is the electric potential. For this quasi-neutral plasma with cold ions, we can evaluate the charge density as
\begin{equation}
	\rho(z,t) = e\left(n_0 - \int \text{d}v f\right),
\end{equation}	
where $n_0$ is the mean number density of both species.

Plasmas with near-Maxwellian velocity distributions are of particular interest for magnetic fusion applications.  Therefore, we separate the distribution function into a Maxwellian mean and time-dependent perturbations:
\begin{equation}\label{eq:split-f}
	f(z,v,t) = F_0(v) + g(z,v,t),
\end{equation}
where 
\begin{equation}
	F_0(v) = \left(\frac{n_0}{v_{te}}\right) \frac{1}{\sqrt{2\upi}} \text{e}^{-v^2/2v_{te}^2},
\end{equation}
is the one-dimensional Maxwellian velocity distribution for the electrons, and $g(z,v,t)$ is the fluctuating part of the distribution function. We have defined the electron thermal speed as $v_{te} = \sqrt{T_e / m_e}$, where $T_e$ is the temperature of the electrons in units of energy.  The Vlasov equation supports longitudinal Langmuir waves, whose dynamics occur on time scales on the order of the plasma frequency,
\begin{equation}
	\omega_{pe} \equiv \sqrt{\frac{4\upi n_0 e^2}{m_e}}.
\end{equation}
The characteristic length-scale for those oscillations is the Debye length,
\begin{equation}
\lambda_D \equiv \sqrt{\frac{ T_e}{4 \upi n_0 e^2}}.
\end{equation}

To support numerical integration of this system of equations, we normalize the variables into dimensionless forms, denoted with primes:

\begin{equation}\label{eq:normalizations}
\begin{split}
	&t \equiv \frac{1}{\omega_{pe}}t', \quad v \equiv v_{te} v', \quad z \equiv \lambda_D z', \quad E \equiv \frac{T_e}{e\lambda_D} E', \quad f \equiv F_0 + g  \equiv \frac{n_{0}}{v_t}(F'_M + g'), \\
	&\rho \equiv n_0 e \rho', \quad \Phi \equiv \frac{T_e}{e}\Phi', \quad F_M' \equiv \frac{1}{\sqrt{2 \upi}} \text{e}^{-v'^2/2}, \quad \int \text{d} v' F_M'= 1.
\end{split}
\end{equation}
After dropping the primes for legibility, we can now write the model equations in dimensionless form: 

\begin{equation}
	\pdt{f} + v \pd{f}{z}  - E\pdv{f} = 0.
\end{equation}
\begin{equation}
	\pd{E}{z} = 1 - \int \text{d}v f.
\end{equation}
Finally, after splitting the distribution function into its mean and fluctuating components, we arrive at the normalized evolution equation for the fluctuating part of the electron distribution function:
\begin{equation}
	\label{eq:genvlasov}
	\pdt{g} + v\pd{g}{z} + vEF_M - E\pd{g}{v} = 0,
\end{equation}
\begin{equation}
	\label{eq:electricfield}
	\pd{E}{z} = -\int \text{d} v \, g.
\end{equation}

\section{Pseudo-spectral Methods}\label{sec:pseudo-spectral}
 We solve the Vlasov-Poisson system (\ref{eq:genvlasov}) - (\ref{eq:electricfield}) using an orthonormal spectral basis by projecting the velocity-space dependence onto a Hermite polynomial basis and the configuration-space dependence onto Fourier harmonics.  
 We choose this approach because pseudo-spectral methods are often more efficient than pure finite-difference algorithms \citep{Boyd00}, and as mentioned in the introduction, there is an extensive history of the application of the Fourier-Hermite basis to solve problems in kinetic plasma physics. 

\subsection{Hermite Moment Formalism}
To discretize velocity space, we impose a Hermite basis representation, using notation from \citet{Mandell18}.
Including both the Maxwellian weight and the normalization factors, the Hermite basis expansion ($\varphi^m(v)$) and projection ($\varphi_m(v)$) functions are 
\begin{equation}\label{eq:hermite_basis}
	\varphi^{m}(v) = \frac{\text{He}_m(v)}{\sqrt{2\upi m!}}e^{-v^2/2} , \quad \varphi_m(v) = \frac{\text{He}_m(v)}{\sqrt{m!}},
\end{equation}
where 
\begin{equation}
	\text{He}_m(v) = (-1)^m e^{v^2/2}\frac{\text{d}^m}{\text{d}v^m} e^{-v^2/2}
\end{equation}
are the probabilists' Hermite polynomials \citep{AbramowitzStegun}.  Note that these functions are orthonormal,
\begin{equation}\label{eq:orthonormality}
	\int_{-\infty}^{\infty} \text{d}v \varphi^m(v)\varphi_{m'}(v) = \delta_{m,m'},
\end{equation}
and that the background Maxwellian is identical to the $m=0$ basis element:
\begin{equation}
	\varphi^0 = \frac{1}{\sqrt{2\upi}}e^{-v^2/2} = F_M.
\end{equation}  This property makes the Hermite moment representation a natural choice for near-Maxwellian velocity distributions.  We project the fluctuating portion of the distribution function onto the Hermite basis using
\begin{equation}
	\label{eq:define_g}
	g(z,v,t) = \sum_{m=0}^{\infty}\varphi^m G_m(z,t), \quad G_m(z,t) \equiv \int_{-\infty}^{\infty} \text{d}v \varphi_m g(z,v,t). 
\end{equation}

\subsection{Vlasov Equation in the Hermite basis}\label{sec:VP_Hermite}
We seek an evolution equation for the amplitudes, $G_m$, of the Hermite expansion functions.  To derive this moment hierarchy, we project (\ref{eq:genvlasov}) onto the Hermite basis term-by-term.  The first term is the most straightforward:
\begin{equation}
\begin{split}
	\int_{-\infty}^{\infty} \text{d}v \varphi_m \pd{g}{t} = \pdt{G_m}.
\end{split}
\end{equation}
We then project the linear ballistic term, $v\left( \partial g  / \partial z\right )$, onto the basis:
\begin{equation}\label{eq:lin_recurrence}
	v\pd{g}{z} = \pd{ }{z}\sum_{m=0}^{\infty} v \varphi^m G_m = \pd{ }{z}\sum_{m=0}^{\infty} \left(\sqrt{m+1}\varphi^{m+1} + \sqrt{m}\varphi^{m-1} \right) G_m,
\end{equation}
where we used a recurrence relation for the Hermite basis functions to remove the explicit dependence on $v$ \citep{AbramowitzStegun}:
\begin{equation}
	\sqrt{m+1}\varphi^{m+1} = v\varphi^m - \sqrt{m}\varphi^{m-1}.
\end{equation}
Applying the projection function to (\ref{eq:lin_recurrence}), and using (\ref{eq:orthonormality}) yields

\begin{equation}
\begin{split}
	\int_{-\infty}^{\infty} \text{d}v \varphi_{m} v\pd{g}{z} = \pd{}{z} \left(\sqrt{m}G_{m-1} + \sqrt{m+1}G_{m+1}\right).
\end{split}
\end{equation}
The linear portion of the electric field drive term contains the $m=1$ expansion function, so applying the projection function to this term yields
\begin{equation}
	\int_{-\infty}^{\infty} \text{d}v \varphi_{m} vEF_M = E\int_{-\infty}^{\infty} \text{d}v \varphi_{m} \varphi^1 = E\delta_{m,1}.
\end{equation}
The final term to consider is the nonlinear term, $-E(\partial g / \partial v)$.
Combining a second recurrence relation \citep{AbramowitzStegun},
\begin{equation}
	\frac{\text{d}}{\text{d}v}\text{He}_m(v) = v\text{He}_m(v) - \text{He}_{m+1}(v),
\end{equation}
with the definition of the Hermite basis functions $\varphi^m$ (\ref{eq:hermite_basis}), presents a useful new identity:
\begin{equation}
	\pd{\varphi^m}{v} = -\sqrt{m+1}\varphi^{m+1}.
\end{equation}
The expansion of the nonlinear term in the Hermite basis is then
\begin{equation}
	-E\pd{g}{v} = -E \sum_m \pd{\varphi^m}{v} G_m = E \sum_m \sqrt{m+1}\varphi^{m+1}G_m.
\end{equation}
After projecting this term onto the basis, 
\begin{equation}
	\begin{split}
		\int_{-\infty}^{\infty} \text{d}v \varphi_{m} (E\sum_{m'} \sqrt{m'+1}\varphi^{m'+1}G_{m'}) &= \sum_{m'} E\sqrt{m'+1}\delta_{m,m'+1}G_{m'} \\ &= E\sqrt{m}G_{m-1},
	\end{split}
\end{equation}
we can collect all of the terms to recover the general form of the equation for $G_m$:
\begin{equation}\label{eq:general}
	\pdt{G_m} + \pd{}{z}\left(\sqrt{m}G_{m-1} + \sqrt{m+1}G_{m+1}\right) + E \sqrt{m}G_{m-1} + E \delta_{m,1} = 0.
\end{equation}

\subsection{Pseudo-spectral in Fourier Space}
Next, we discretize space with a Fourier basis. The Fourier transform and inverse Fourier transform operators for a spatial grid of $N$ points are:
\begin{equation}\label{eq:fourier}
	\mathcal{F}[g] = \frac{1}{N} \sum_{z}g \text{e}^{-\text{i}kz}\quad \mathcal{F}^{-1}[G_k] = \sum_k G_k \text{e}^{\text{i}kz},
\end{equation}
where the Fourier wavenumbers  are given by $k = 2\upi j / N$, with $-N/2 \leq j < N/2$ and $j \in \mathbb{Z}$.  The Fourier convolution theorem,
\begin{equation}
	\mathcal{F}[f g] = \sum_{k'}F_k G_{k-k'},
\end{equation}
transforms the nonlinear term into a coupled sum over all wavenumbers.  This coupling introduces additional computational overhead. We mitigate that problem by using a pseudo-spectral method \citep{Boyd00}, evaluating the nonlinear term on the original configuration space and projecting the result back onto the space of Fourier wavenumbers.  Applying the Fourier transform, \eqref{eq:fourier}, to \eqref{eq:general} and then applying the Fourier convolution theorem yields
\begin{multline}
	\label{eq:gen_form}
	\pdt{G_{m,k}} + \text{i}k\left(\sqrt{m}G_{m-1,k} + \sqrt{m+1}G_{m+1,k}\right) + \sqrt{m}\mathcal{F}\left[\mathcal{F}^{-1}[E_k] \cdot \mathcal{F}^{-1}[G_{m-1,k}]\right] \\
	 = - E_k \delta_{m,1}.
\end{multline}

Simulating this collisionless system with finite resolution without using any regularization is numerically infeasible, as the dynamics can generate structures in phase space at arbitrarily small scales \citep{ParkerDellar, Loureiro16, Mandell24}.  To mitigate that effect, we introduce numerical regularization through hyperdiffusion $(-\nu_k k^4)$ and hypercollision $(-\nu_m m^4)$ operators:
\begin{equation}
\label{eq:hyper_regularization}
	H_{mk} = \begin{cases}
		0 & \text{if $m \leq 2$} \\
		-\nu_k k^4 -\nu_m m^4 & \text{if $m>2$},
	\end{cases}
\end{equation}
where $\nu_k$ and $\nu_m$ are parameters which control the strengths of the hyperdiffusion and hypercollisions, respectively.  Note that we do not apply this regularization to the $m=$ 0, 1, and 2 equations, as they encode conservation of particle number, momentum, and energy, respectively.

We solve the system for a finite number of moments $M$.
Defining compact notation for the nonlinear operator,
\begin{equation}\label{eq:nonlinear_operator}
	\mathcal{N}[E_k,G_{m-1,k}] = \sqrt{m}\fft{\ifft{E_k}\cdot \ifft{G_{m-1,k}}},
\end{equation}
and accounting for the special cases of the $m=0$, $m=1$, and $m=M-1$ equations, we have derived the projection of the Vlasov equation onto the Fourier-Hermite basis:
\begin{alignat}{2}
	\label{eq:full_NL_VP_spectral_start}
	&\pdt{G_{0,k}} &&+ \text{i}kG_{1,k} = 0, \\	
	&\pdt{G_{1,k}} &&+ \text{i}k\left(G_{0,k} + \sqrt{2}G_{2,k}\right) + \mathcal{N}[E_k,G_{0,k}] + E_k = 0, \\
	&\pdt{G_{m,k}} &&+ \text{i}k\left(\sqrt{m}G_{m-1,k} + \sqrt{m+1}G_{m+1,k}\right) + \mathcal{N}[E_k,G_{m-1,k}] \nonumber \\
	&\phantom{} &&= H_{m,k} G_{m,k}, \quad 2 \leq m < M-1,  
	\label{eq:gen_form_hyper}
	\\
	&\pdt{G_{M-1,k}} &&+ \text{i}k\sqrt{M-1}G_{M-2,k} + \mathcal{N}[E_k,G_{M-2,k}] =  H_{M-1,k} G_{M-1,k}
	\label{eq:full_NL_VP_spectral_end}.
\end{alignat}

\subsection{Poisson's Equation in the Fourier-Hermite basis}

To complete the system, we project Poisson's equation onto the Fourier-Hermite basis and solve for the electric field, $E_k$.  First, we solve for the charge density in the Hermite basis:
\begin{equation}
		\rho = -\int_v \text{d}v g  = -\int_v \text{d}v \varphi_0 \sum_m \varphi^m G_m = -G_0.
\end{equation}
Next, we insert this expression for the normalized charge density into Poisson's equation (\ref{eq:electricfield}) and apply the Fourier transform (\ref{eq:fourier}) to calculate the electric field as a function of the $m=0$ density moment in the Fourier-Hermite basis:
\begin{equation}
	\label{eq:field_solve}
	E_k = \frac{\text{i}G_{0,k}}{k}.
\end{equation}
We require that the $k=0$ mode is zero and stationary, corresponding to zero mean electric field.

\subsection{The Closure Problem}\label{sec:closure}
A continuous representation formally requires an infinite number of Hermite moments, but we must discretize the system with a finite number of moments, $M$.  
Observe that in the Fourier-Hermite basis, the parallel streaming term in the convective derivative, $v (\partial g / \partial z)$, becomes $\text{i}k\left(\sqrt{m}G_{m-1, k} + \sqrt{m+1}G_{m+1, k}\right)$ for general $m$.  This term couples nearest-neighbors in Hermite space and forms a hierarchy of equations that must be closed.  
This closure problem is analogous to the classic closure problem encountered when taking fluid moments of the kinetic equation (\ref{eq:vlasov_unnormalized}). 
The resulting fluid equations for density, momentum, energy, and higher order moments exhibit coupling between moment $m$ and moments $m+1$ and $m-1$.
This same feature occurs in the Hermite representation of velocity space, and from (\ref{eq:define_g}) and (\ref{eq:hermite_basis}), the Hermite moment basis is obtained through a similar process of multiplying the distribution function by increasing powers of $v$ via increasing orders of the Hermite polynomials and then integrating over velocity space.

The simplest model of closure is by truncation, which we use in the direct numerical solution by omitting the $G_{m+1,k}$ term in the evolution equation for the final moment, $G_{M-1,k}$.  Our objective is to introduce a dynamical closure model to close the hierarchy at a moment number $m=m_c, \, m_c < M-1$, while still achieving accurate predictions of the Fourier and Hermite spectra.  We use a reservoir recurrent neural network as the closure model, which we describe in more detail in Section \ref{sec:reservoir}\,.

\section{Machine-Learning Closure: Reservoir Computing}
\label{sec:reservoir}
Inspecting the evolution equation for $G_{m,k}$ (\ref{eq:general}) reveals that the coupling between moment $m$ and moment $m+1$ only explicitly appears as a linear term in the equation, acting on each wavenumber $k$ independently.  As a result, we introduce a unique and independent Hermite closure model for each $k$ in the system, for a total of $N_k$ reservoirs.  
We use methods similar to \citet{PathakPRL} to design the reservoirs. 

\begin{figure}
\centering
\includegraphics[scale=0.4]{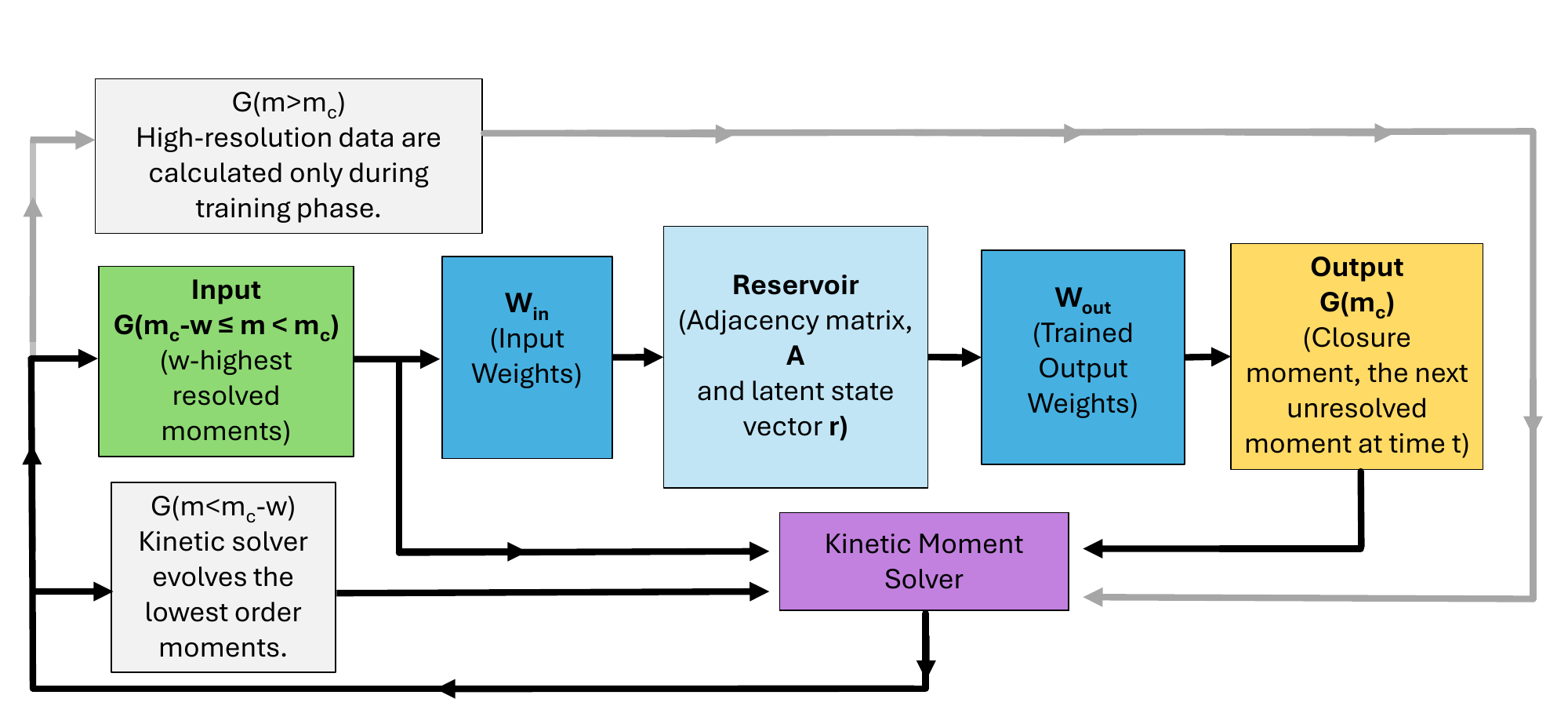}
\caption{Diagram of the reservoir computing closure model, integrated with the kinetic moment solver.}
\label{fig:closure_diagram}
\end{figure}

Each reservoir is a network of $D_{r}$ nodes defined by a latent state vector $ \boldsymbol{r}_k(t) \, \in \mathbb{R}^{D_r}$, where each element of $\boldsymbol{r}_k(t)$ represents the state of one node at time $t$.  
The structure of the reservoir computing model is presented in Figure \ref{fig:closure_diagram}\,.  
The connections between nodes are defined by a $D_{r} \times D_{r}$ sparse adjacency matrix, $\mathsfbi{A}$, where exactly $\kappa$ nonzero elements per row are initialized to random values drawn from the uniform distribution on [0,1] and scaled by the appropriate factors to set the largest eigenvalue of $\mathsfbi{A}$ equal to a specified value, $\rho_{sp}$. 
The magnitude of the largest eigenvalue of a reservoir's adjacency matrix has been demonstrated to significantly impact the accuracy of the reservoir's predictions \citep{Jiang19}.  
The runtime is separated into two phases, training and prediction. 
In the training phase, the inputs are derived from a high-resolution direct numerical solution (DNS).  
In the subsequent prediction phase, the inputs are derived from a low-resolution simulation with the output of our reservoir used to compute the closure.  
The input to a given reservoir at each time step is a vector $\boldsymbol{u}_k(t)$ of dimension $2w$, where the elements of $\boldsymbol{u}$ are the real and imaginary parts of the final $w$ Hermite moments before $m_c$:
\begin{equation}
	\boldsymbol{u}(t) = \left[G_{m_c-w,k}(t-w\Updelta t), \, G_{m_c-(w-1),k}(t-(w-1)\Updelta t), \,  \ldots  G_{m_c-1,k}(t-\Updelta t)\right].
\end{equation}
In both phases, the inputs form a diagonal in $m$ and $t$ of Hermite moments, and this structure is inspired by the phenomenon of linear Landau damping in the linearized Vlasov-Poisson system, where energy cascades from low $m$ to high $m$ in time.  The input layer is a $D_{r} \times 2w$ matrix $\mathsfbi{W}_{in}$, with its elements drawn from the uniform distribution on $[-\sigma,\sigma]$, where $\sigma$ is a parameter. Critically, $\mathsfbi{A}$ and $\mathsfbi{W}_{in}$ remain fixed after initialization.  Only the output layer, $\mathsfbi{W}_{out}$, is trained.  The reservoir state vector evolves with the equation
\begin{equation}
	\label{eq:reservoir_evolution}
	 \boldsymbol{r}_k(t + \Updelta t) = \tanh[\mathsfbi{A} \boldsymbol{r}_k(t) + \mathsfbi{W}_{in}\boldsymbol{u}(t)],
\end{equation}
where the hyperbolic tangent activation function operates element-wise on the vector argument.  

The elements of the reservoir-to-output coupling, $\mathsfbi{W}_{\text{out}}$, are parameters that must be tuned to train the reservoir. We introduce a hyperparameter $b$ to exclude data from initial transients from the reservoir training process.  This hyperparameter serves as a buffer of time steps between the initialization time and the data assigned for training $\mathsfbi{W}_{\text{out}}$. During the training phase, defined by the times $t$ where $b \Updelta t \leq t \leq T$, the reservoir output is trained to approximate the closure moment, $G_{m_c,k}$. 

The most straightforward choice for the closure evaluation would be the linear readout operation, $\mathsfbi{W}_{out} \boldsymbol{r}_k$.  
Empirically, that operation does not permit reservoirs to predict the dynamics of several other nonlinear systems, including the Lorenz system \citep{Lu17, Chattopadhyay20, Pyle21}, Kuramoto-Sivashinsky equation \citep{PathakPRL}, and the global atmospheric system \citep{Arcomano20}. 
As \citet{Lu17} and \citet{PathakPRL} describe, the hyperbolic tangent operation in the evolution equation for the reservoir state (\ref{eq:reservoir_evolution}) introduces an odd-parity symmetry to the reservoir state vector $\boldsymbol{r}$.
As in the Kuramoto-Sivashinsky equation and the Lorenz systems studied by these authors, the quadratic nonlinear term in the Vlasov equation breaks odd-parity symmetry.
In other words, the general evolution equation for the Hermite moments (\ref{eq:gen_form}) is \textit{not} invariant under the transformation $G_{m,k} \rightarrow -G_{m,k}$, while the evolution equation for the reservoir state (\ref{eq:reservoir_evolution}) \textit{is} invariant under the transformations $\boldsymbol{u} \rightarrow -\boldsymbol{u}$ and $\boldsymbol{r} \rightarrow -\boldsymbol{r}$.
Following the methods of \citet{PathakPRL} and explored in more detail by  \citet{Chattopadhyay20} and \citet{Pyle21}, we resolve this conflict by applying a symmetry-breaking transformation ($\boldsymbol{r}_k \rightarrow \boldsymbol{r}_k^*$) to the reservoir latent state before readout:

\begin{equation}
	\label{eq:symmetry_break}
	r_{i,k}^*(t) = 
	\begin{cases}
		r_{i,k}(t), \quad \text{even  } i \\
		r_{i,k}^2(t), \quad \text{odd  } i.
	\end{cases}
\end{equation}
Because the reservoir output, $\mathsfbi{W}_{out}\boldsymbol{r}^*(t)$, is a linear operation on $\boldsymbol{r}^*$, we can train the weights of $\mathsfbi{W}_{out}$ using linear regression and avoid the computational expense of backpropagation.  To mitigate potential overfitting, we introduce a regularization parameter, $\beta$, in the method of Tikhonov-regularized linear regression.  The training problem is then to find the values of $\mathsfbi{W}_{out}$ which minimize 
\begin{equation}
	\sum_{j = b}^{T/\Updelta t}||G_{m_c,k}(j\Updelta t) - \mathsfbi{W}_{\text{out}} \boldsymbol{r}_k^*(j\Updelta t)||^{2} + \beta ||\mathsfbi{W}_{out}||^{2}.
\end{equation}
Explicitly, the closed-form solution that solves this linear regression problem is
\begin{equation}
	\label{eq:closed_form_regression}
	\mathsfbi{W}_{out} = \mathsfbi{G}_{m_c,k} \mathsfbi{R}_k^{*T}(\mathsfbi{R}_k^*\mathsfbi{R}_k^{*T} + \beta \mathsfbi{I})^{-1},
\end{equation}
where $\mathsfbi{G}_{m_c,k}$ is a $2 \times (T/\Updelta t - b)$ matrix consisting of the time series of the real and imaginary parts of the closure moment, $G_{m_c,k}$, $\mathsfbi{R}_k^*$ is a $D_r \times (T/\Updelta t- b)$ matrix consisting of the time series of the modified state vector $\boldsymbol{r}_k^*(t)$, $\mathsfbi{R}_k^{*T}$ is the transpose of $\mathsfbi{R}_k^*$, and $\mathsfbi{I}$ is the identity matrix.  

\subsection{Closure Integration}
\label{sec:closure_int}

Once the weights of $\mathsfbi{W}_{out}$ are trained, the algorithm switches to the prediction phase, and the system does not evolve moments beyond $m = m_c$.  Instead, the output of the reservoir is fed back into the system of equations as a closure for the moment hierarchy.  To make the discussion more concrete, we introduce a time integration operator $\mathcal{D}$, such that $\mathcal{D}[\mathsfbi{G}(t)] = \mathsfbi{G}(t + \Updelta t)$, where $\mathsfbi{G}(t)$ is the $M \times N_k$ matrix of Fourier-Hermite amplitudes representing the state of the system at time $t$.  $\mathcal{D}$ is a generic operator, and its implementation details are flexible.  For the results presented in this paper, we implement $\mathcal{D}$ using the Python libraries \texttt{NumPy} \citep{Harris20} and \texttt{SciPy} \citep{Virtanen20} with the classic fourth-order explicit Runge-Kutta method (RK4) and the third-order Strong Stability-Preserving Runge-Kutta method (SSPRK3) as detailed by \citet{Durran}.

When we introduce the closure models, the time integration becomes a two-step process.  For each wavenumber $k$, we use the time integration operator $\mathcal{D}$, to calculate the resolved moments $\boldsymbol{G}_{m<m_c,k}$:
\begin{equation}
	\mathsfbi{G}_{m<m_c,k}(t + \Updelta t) = \mathcal{D}[\mathsfbi{G}_{m\leq m_c}(t)]_k.
\end{equation}
We then use a reservoir to predict the closure moment $G_{m_c,k}$:
\begin{equation}
	\label{eq:closure_prediction}
	G_{m_c, k}  (t + \Updelta t) = \mathsfbi{W}_{out} \boldsymbol{r}_k^*(t + \Updelta t).
\end{equation}

\subsection{Computational Complexity}
Though performance profiling metrics for numerical solutions of partial differential equations are highly dependent on computational hardware architectures and algorithm implementation choices, we can calculate the computational complexity of the DNS and the ML closure for the cases in this paper.
Both methods have an identical cost to evolve the lowest-order moments. 
Thus, we only report the cost of evolving moments $m_c \leq m < M$ for the DNS and the closure moment $m_c$ for the ML method.
While the time-integration operator $\mathcal{D}$ is generic, we will remove a layer of abstraction by calculating the result for the SSPRK3 algorithm.  
For simplicity, we will neglect the additional cost of operations on complex numbers relative to those same operations on real numbers, with the understanding that this will underestimate the cost of the DNS in comparison to the ML closure.
The cost of transcendental functions including $\tanh$ depends on instruction sets.
This estimate treats them as equivalent in cost to multiplication, with the understanding that this underestimates the cost of updating the reservoir state.
We also neglect the cost of memory access and copy operations for both methods.

\subsubsection{Complexity of DNS}
First, let us introduce a new parameter $M' = M - m_c$ that represents the additional number of moments that the DNS must evolve (\ref{eq:gen_form_hyper}) in comparison to the ML closure model.
Our choice of time-integration operator, SSPRK3, is a three-stage algorithm. 
Each stage includes an evaluation of the electric field, nonlinear term, linear streaming term, and hypercollisions.
The electric field is calculated by (\ref{eq:field_solve}).  
The vector $(\text{i} / k)$ can be stored in memory during initialization, so the electric field calculation requires $N_k$ multiplication operations.
The nonlinear term is evaluated with a pseudo-spectral method using (\ref{eq:nonlinear_operator}).
Using the $\mathcal{O}(N \log N)$ scaling of the fast Fourier Transform (FFT) algorithm \citep{Cooley65}, the total computational complexity of the nonlinear term is $3 M' \mathcal{O}(N_k \log N_k)$.
If the vector $(\text{i} k)$ is also stored during initialization, then the linear streaming term requires $6M'N_k$ operations.
The dissipation matrix in \ref{eq:hyper_regularization} can be stored, and the cost of the regularization term is $2M'N_k$.
Finally, we include the operations needed to combine the results of each stage into the state of the system at the next time step.
For SSPRK3, this requires $12M'N_k$ operations.
The total cost of DNS for each time step is then
\begin{equation}
	\label{eq:DNS_cost}
	\text{Cost}_{\text{DNS}} = 3M'\mathcal{O}(N_k \log N_k) + 20M'N_k + N_k.
\end{equation}

\subsubsection{Complexity of ML Closure}
The ML closure evaluation has two stages, updating the latent state of each reservoir and evaluating the closure moment.
There are a total of $N_k$ reservoirs.
The state of each reservoir, $\boldsymbol{r}_k$, updates at every time step using (\ref{eq:reservoir_evolution}).
This requires the sum of two matrix-vector products.
The product $\mathsfbi{W}_{in} \boldsymbol{u}$ has a cost of $4D_rw$ operations.
The other product $\mathsfbi{A} \boldsymbol{r}_k$ involves a sparse matrix.  
Though the cost of this product is implementation-specific, we will count operations on nonzero elements, resulting in a cost of $2 D_r \kappa$ operations, where $\kappa < D_r$.
Then, the $\tanh$ operation is applied to the sum of these products.
The total cost of the reservoir update is then $2D_r[\kappa + 2w + 1]$.
The symmetry-breaking transformation (\ref{eq:symmetry_break}) is applied to the state vector, which requires $D_r / 2$ multiplication operations.
Finally, each reservoir predicts a Hermite closure using (\ref{eq:closure_prediction}), which uses $4D_r$ operations.
The ML closure is linear in $k$.
In total, the cost calculating the ML closure for each time step is then
\begin{equation}
	\label{eq:ML_cost}
	\text{Cost}_{\text{ML}} = N_k D_r [8w + 4\kappa + 13/2].
\end{equation}
After eliminating the common factor of $N_k$ from the cost of the DNS and the ML closure, we find that as $N_k$ increases, the dominant scaling for the DNS is $\mathcal{O}(M' \log N_k)$.  In contrast, the cost of the ML closure scales linearly with the size of each reservoir $D_r$.
For sufficiently large $(M' \log N_k) / (D_r w)$, the reservoir closure will be faster than the DNS.

\section{Closure Results}
\label{sec:results}

\subsection{Linearized Vlasov-Poisson}
\label{sec:linearized_VP_results}

\begin{figure}
	\begin{minipage}{0.45\textwidth}
		\includegraphics[scale=0.5]{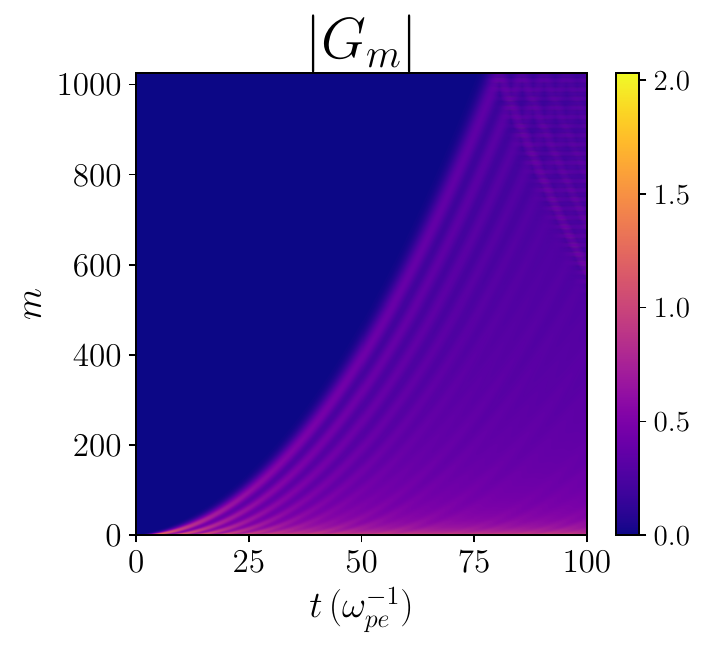}
		\caption{
			\label{fig:linearized_forced}
			High-velocity-resolution ($M=1025$) baseline time series of Hermite amplitudes for the driven, linearized Vlasov-Poisson system.  Energy is injected as a density perturbation in the $m=0$ Hermite moment and advects to higher moments through linear Landau damping.  No hypercollisional regularization is applied to this example, and hundreds of Hermite moments are required to prevent numerical reflection at the high-$m$ boundary from impacting the low-$m$ spectrum.}
	\end{minipage}
	\hfill
	\begin{minipage}{0.45\textwidth}
		\includegraphics[scale=0.43]{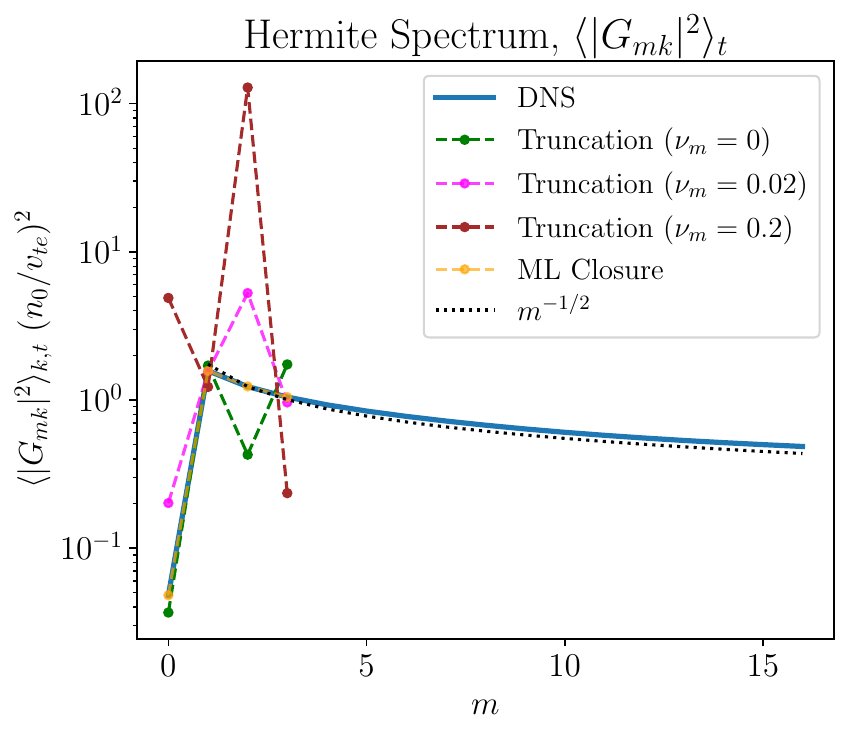}
		\caption{
			\label{fig:linear_spectrum}
			Time-averaged Hermite spectra for the ML closure model compared to the high-velocity-resolution ($M=1025$) baseline, closure-by-truncation with three different coefficients ($\nu_m$) for hypercollisional regularization, and the theoretical $m^{-1/2}$ scaling. The ML closure model shows strong agreement with the baseline simulation, while no value of $\nu_m$ produces an accurate spectrum.}
	\end{minipage}
\end{figure}

For our first test of the ML closure, we confirm that the model accurately solves the linear limit of the Vlasov-Poisson system. 
First, we linearize the Vlasov equation (\ref{eq:genvlasov}), dropping the nonlinear term:
\begin{equation}
	\pdt{g} + v\pd{g}{z} +vEF_M = 0.
\end{equation}
In Fourier space, this allows us to restrict our focus to a single wavenumber, as the coupling between wavenumbers occurs only in the nonlinear term.
We simulate an approximately steady-state solution by driving the system, as explored by \citet{Kanekar15}. We inject energy into the $m=0$ moment, and it cascades down to finer scales through Landau damping.  
We incorporate these features into the projection of this equation onto the Fourier-Hermite basis using the results of Section \ref{sec:pseudo-spectral}\,:
\begin{equation}
	\label{eq:driven_linear}
	\pdt{G_{m,k}} + \text{i}k\left(\sqrt{m}G_{m-1,k} + \sqrt{m+1}G_{m+1,k}\right) = \chi(t)\delta_{m,2} - E_k \delta_{m,1}, 
\end{equation}
where $\chi(t)$ is a forcing coefficient independently drawn at each time step from the uniform distribution on $[0,1)$. 
The electric field $E_k$ is given by (\ref{eq:field_solve}). 
The purpose of this mechanism for energy injection is to achieve a statistical steady-state solution.  
Our primary systems of interest exhibit nonlinear chaos and typically settle into time-dependent steady states.  
While constant forcing would also achieve a steady-state solution, a time-dependent forcing mechanism more closely resembles typical problems encountered in plasmas.
Other injection mechanisms, including driving the $m=2$ moment or applying a stochastic external electric field yield similar performance for the closure model.
Figure \ref{fig:linearized_forced} depicts a high-resolution ($M=1025$) baseline simulation of (\ref{eq:driven_linear}) with $k=0.4$.  

When no hypercollisional regularization is applied, the cascade of free energy numerically reflects at the high-$m$ boundary.
In the simulation in Figure \ref{fig:linearized_forced}, this reflection occurs around $t=80$.
This phenomenon always occurs at a finite time for a given Hermite resolution ($M$), meaning that extending the duration of a simulation where the low-order moments are valid requires increasing the Hermite resolution.
Additionally, the rate at which energy advects from low-$m$ to high-$m$ increases with Fourier wavenumber $k$ \citep{ParkerDellar}.  
Therefore, though this linearized simulation with $k=0.4$ is feasible with high velocity resolution, more Hermite resolution would be required for a nonlinear simulation with moderate spatial resolution.
Some form of artificial dissipation is necessary for baseline simulations of higher-dimensional systems to be computationally feasible.
We therefore apply hypercollisional regularization to the nonlinear form of the Vlasov-Poisson system in Sections \ref{sec:weakly_NL_VP} and \ref{sec:nonlinear_VP}\,.

We then train a reservoir to predict the Hermite closure moment, $G_{m_c,k}$, where we evaluate the closure at $m_c = 3$. 
The choice of $m_c=3$ is motivated by the tradition of heat flux closures in fluid theory and our desire for a closure that allows the conservation laws in the $m \in [0,1,2]$ equations to remain intact.
One motivation for a closure is to reduce the resolution required to achieve accurate simulations, so we choose the lowest valid value for $m_c$.
We use the time series of the density, momentum, and temperature moments from the time window $15 \leq t < 25$ as training data.
We find that a requirement is that enough time has passed to allow energy to cascade into $m=m_c$.  We report results training on data from the time window $15 < t \leq 25$, but the reservoir supports accurate spectra for other training windows as well.
Figure \ref{fig:linear_spectrum} depicts the excellent agreement between the low-order spectra calculated by the high-resolution baseline and the ML closure model. 
The mean-squared error in the ML closure spectrum is $2.25 \times 10^{-5}$.
The ML closure shows favorable performance in comparison to a na\"{i}ve closure by truncation, where no hypercollisional regularization is applied. 
It also outperforms a closure by truncation where the hypercollisional term, $-\nu_m m^4 G_{m,k}$, is introduced to the right-hand side of (\ref{eq:driven_linear}) to mitigate numerical reflection at the high-$m$ boundary.
As in (\ref{eq:hyper_regularization}), this term is only applied for $m>2$ to preserve the low-order conservation laws. 
For an initial comparison, $\nu_m$ is set to $2 \times 10^{-2}$ to maintain finite dissipation at $m_c = 3$, where $\nu_m m_c^4$ is order one.
For reference, we also compare to the theoretical scaling $G_m \sim m^{-1/2}$ \citep{Zocco11, Kanekar15}.

The reservoir hyperparameters used to construct the model are spectral radius $\rho_{sp}=0.6$, adjacency matrix degree of three, Tikhonov regularization parameter $\beta = 10^{-9}$, input scaling parameter $\sigma = 0.5$, and six nodes in the reservoir.  We found these hyperparameters by comparing mean-squared errors in spectra.  An unguided scan led to a reduction in error from $1.75 \times 10^{-7}$ to $2.26 \times 10^{-8}$.  No systematic optimization was required, though some choices of parameters led to numerically unstable time integration of (\ref{eq:driven_linear}). 
We used the same procedure to select the hyperparameters in both the weakly and strongly nonlinear cases presented in Sections  \ref{sec:weakly_NL_VP} and \ref{sec:nonlinear_VP}, respectively.  
The spectral radius $\rho_{sp}$ of the adjacency matrix $\mathsfbi{A}$ must be set to $\rho_{sp} < 1$ for stability.
The block-diagonal structure for $\mathsfbi{W}_{in}$ that \citet{Vlachas20} used in their reservoir computing implementation yielded unstable predictions.
We speculate that this may be related to differences in requirements for complex-valued data in comparison to real-valued data, but more analysis would be required to confirm.
For the strongly nonlinear case in Section \ref{sec:nonlinear_VP}\,, using a reservoir smaller than sixty nodes or including the initial transient in the training set also led to unstable solutions.
The computational complexity for the evaluating the ML closure is the same as in \ref{eq:ML_cost}\,, with $N_k=1$. 
The DNS cost for this case is $18M' + 1$, after removing the contributions of the nonlinear term and hypercollisions from (\ref{eq:DNS_cost}).

\subsection{Weakly Nonlinear Vlasov-Poisson}
\label{sec:weakly_NL_VP}
\begin{figure}
	\begin{minipage}{.46\textwidth}
		\centering
		\includegraphics[scale=0.5]{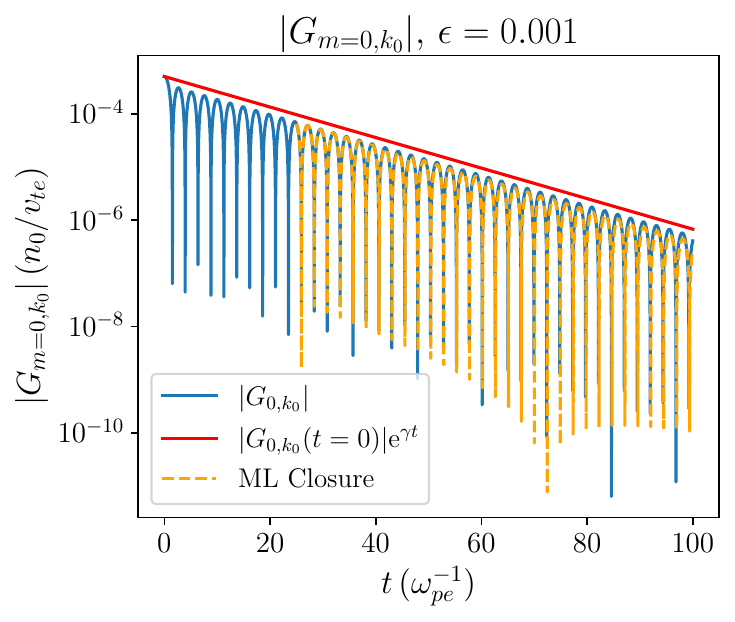}
		\caption{
			\label{fig:linear_landau}
			Comparison between baseline numerical solution, ML closure, and theoretical damping rate of the Fourier-Hermite amplitude for an initial cosine density perturbation.  The low-amplitude perturbation shows strong agreement with the theoretical damping rate. When augmented with the ML closure, the moment solver continues to capture the behavior well at a lower Hermite resolution of $M=4$, as opposed to the $M=17$ baseline.}
			\end{minipage}
	\hfill
	\begin{minipage}{.46\textwidth}
		\centering
		\includegraphics[scale=0.51]{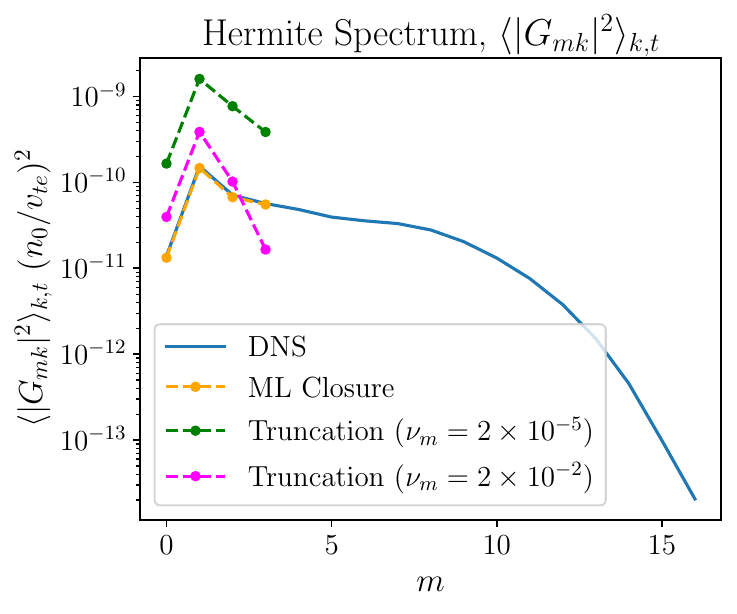}
		\caption{
			\label{fig:linear_reservoir_truncated}
			Hermite spectra of the high-velocity-resolution ($M=17$), truncated ($M=4$) simulations, and ML closure for the initial value problem in Figure \ref{fig:linear_landau}\,.  The spectra are averaged over time and Fourier wavenumber. The ML closure model permits a low-resolution simulation to accurately resolve the Hermite spectrum.}
	\end{minipage}
\end{figure}

\begin{figure}
	\centering
	\includegraphics[scale=0.45]{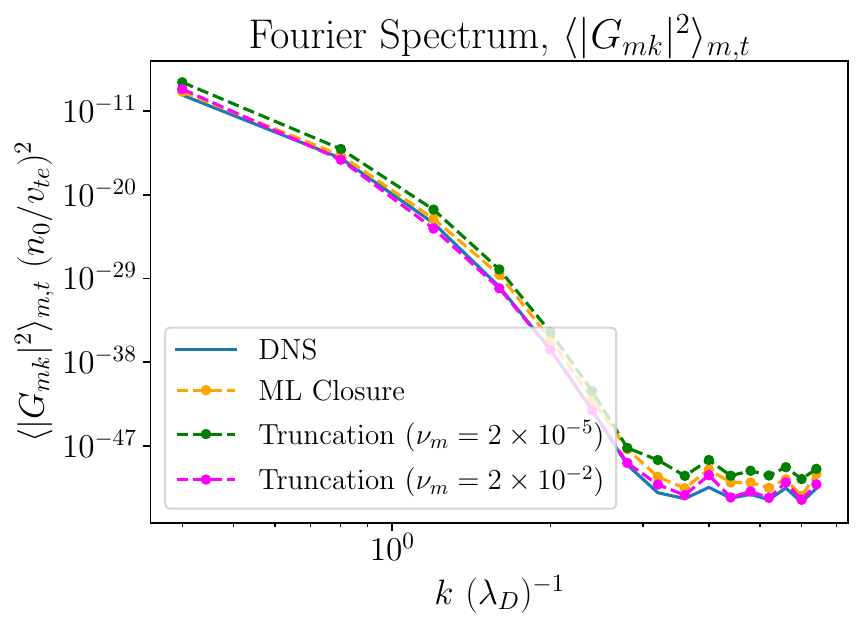}
	\caption{
		\label{fig:linear_w_truncation_fourier}
		Fourier spectra of the simulations in Figure \ref{fig:linear_reservoir_truncated}\,.
		In this weakly nonlinear regime, the majority of the energy remains in the boxscale mode.  Each reduced Hermite resolution model properly captures the rapid energy decay across $k$.
	}
\end{figure}

Next, we reintroduce the nonlinear term, $-E (\partial g / \partial v)$, which permits coupling between wavenumbers. 
This coupling in Fourier space allows energy to flow across spatial scales, in addition to the already-present energy cascade in velocity space.  
This effect is subdominant at low amplitudes, but it becomes more significant as amplitudes increase. 
We first examine the low-amplitude limit in this section before proceeding to a high-amplitude case in Section \ref{sec:nonlinear_VP}\,. In the low-amplitude limit, Landau damping continues to be the dominant effect.  
Here, we solve an initial value problem, where the initial condition is a cosine density perturbation with a Maxwellian velocity component:
\begin{equation}
	g(z,v,t=0) = \epsilon \cos (k_0 z)F_M(v),
\end{equation}
where $\epsilon$ is the initial amplitude of the perturbation normalized by the background density and Maxwellian weight and $k_0$ is the wavenumber of the initial perturbation.  
In the spectral domain, this initial condition has the convenient form
\begin{equation}
	G_{m,k,t=0} = \frac{\epsilon}{2} \delta_{m,0} \left(\delta_{k,k_0} + \delta_{k,-k_0}\right), 
\end{equation}
collapsing the initial system state into a single nonzero value after applying the reality condition.  
To explore the linear limit of the system, we choose $\epsilon = 0.001$, a $0.1 \%$ density perturbation.  
We examine the $k_0 = 0.4$ case that \citet{Brunetti00} explored in their analysis, accounting for a factor of ten difference in normalizations.  
With this low initial amplitude, we choose $N_k=17$ Fourier wavenumbers, including the $k=0$ mode, anticipating that the dominant energy cascade will be linear Landau damping.  
For the high-resolution simulations, we continue to use $M=17$ Hermite moments, including the $m=0$ moment.  
The hypercollisional regularization coefficient is $\nu_m = 5 \times 10^{-5}$, and we set $\nu_k = 0$.  
This value of $\nu_m$ is chosen to maintain finite dissipation at the finest velocity scale in the simulation, such that the magnitude of the coefficient of the hypercollisional term at that scale, $\nu_{m} (M-1)^4$, is order one.
As the Fourier spectrum in Figure \ref{fig:linear_w_truncation_fourier} shows, the flux of energy to high $k$ in this regime is small, so $\nu_k$ is not necessary.
We set the size of the spatial simulation domain to $L=5\upi$ so that $1/k_0$ is the largest wavelength resolved in the system.
Figure \ref{fig:linear_landau} presents a comparison between the baseline, high-Hermite-resolution simulation, the ML closure model, and the theoretical damping rate.  
The time trace for the ML closure method begins at the end of the training phase, $t=25$.
To calculate the damping rate, we use the complex root-finding algorithm of \citet{Carpentier82}, implemented in the generalized dispersion relation solver developed by \citet{Ivanov23}.  
These results demonstrate that with a low initial amplitude, our baseline simulation converges to the expected linear Landau damping rate. Additionally, Figure \ref{fig:linear_landau} demonstrates that the ML closure model preserves the frequency of the wave.
Time traces of the $m=1$ and $m=2$ moments show similar performance and are presented in Appendix \ref{sec:time_traces}\,.

In Figure \ref{fig:linear_reservoir_truncated}\,, we compare the ML closure model, with $m_c = 3$, to both the high-resolution ($M=17$) baseline and two low-resolution simulations without the ML closure, i.e. closure by truncation.  We also compare the Fourier spectra of these simulations in Figure \ref{fig:linear_w_truncation_fourier}\,.  For this problem, we place an independent reservoir at each wavenumber $k$ and train it to learn a Hermite moment closure for that wavenumber.  The ML closure shows strong agreement with the high-resolution baseline, with mean-squared errors of $1.26 \times 10^{-42}$ for the Hermite spectrum and $8.01 \times 10^{-38}$ for the Fourier spectrum.  It successfully captures the Hermite and Fourier spectra of the system and confirms that its predictive capability continues to hold when a small, but nonzero, flux of energy flows between wavenumbers.  To create a low-resolution simulation without the ML closure, we truncate the moment hierarchy at $m_c=3$ and set $G_{m_c+1,k}=0$ in the evolution equation for $G_{m_c,k}$.  
As in Section \ref{sec:linearized_VP_results}\,, we also compare to a low-resolution simulation with increased hypercollisional regularization, with $\nu_m$ set to $2 \times 10^{-2}$ to maintain finite dissipation at the finest resolved scales.  An intuitive explanation for the poor performance of the truncated simulations in Figure \ref{fig:linear_reservoir_truncated} can be described in relation to the general form of the evolution equation for this problem, (\ref{eq:gen_form}). The truncation process removes the $G_{m+1,k}$ term from the final moment equation in the truncated system, eliminating an effective energy sink from that moment.  Without access to the proper dissipation channel of smaller scales in velocity space, the truncated system experiences some energy reflection and pile up in its smallest resolved scales in velocity. In contrast, the ML closure learns the pattern of dissipation through Landau damping and serves as a better boundary condition than truncation.  

We find that the reservoir hyperparameters that result in the best closure performance for this initial condition are spectral radius $\rho_{sp} = 0.6$, Tikhonov regularization parameter $\beta = 10^{-7}$, and $T=25$ normalized time units for training.  For each wavenumber, we set the reservoir input scaling parameter $\sigma$ to normalize the reservoir inputs by the time average of the Hermite amplitude at the moment before the closure:
\begin{equation}
	\sigma_k = \frac{1}{\langle |G_{m_c-1,k,t}| \rangle_t}.
\end{equation}
This preserves the dynamic range of the hyperbolic tangent nonlinear activation function by mitigating the possibility of saturation to $-1$ or 1.
As in Section \ref{sec:linearized_VP_results}\,, we set the degree of the adjacency matrix to three.  Finally, we choose a reservoir size of two nodes per input value, totaling twelve nodes.

\subsection{Strongly Nonlinear Vlasov-Poisson}
\label{sec:nonlinear_VP}

\begin{figure}
	\begin{minipage}{.46\textwidth}
		\centering
		\includegraphics[scale=0.46]{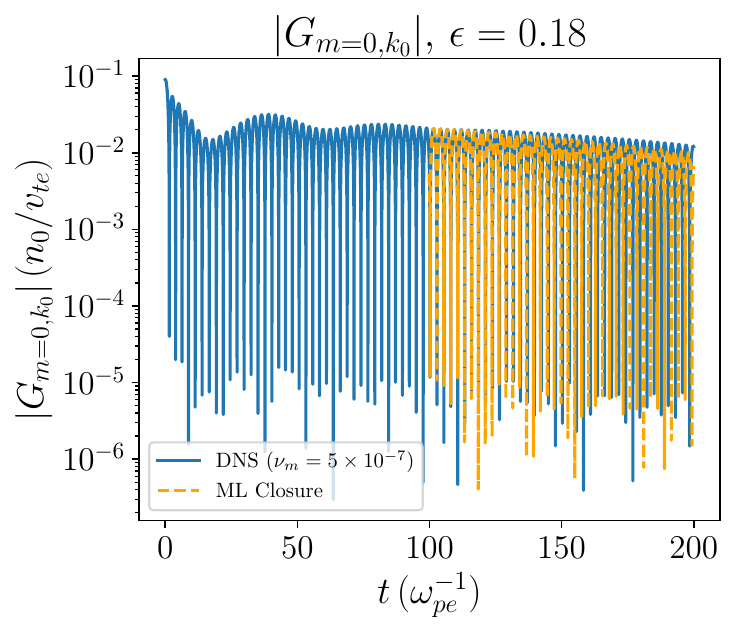}
		\caption{
			\label{fig:NL_baseline}
			Simulated Fourier-Hermite amplitude for a high initial amplitude (18\% of background) cosine density perturbation.  The dynamics exhibit strongly nonlinear behavior, attenuating the damping of the mode.
			As in the low-amplitude case, the ML closure model captures the frequency and amplitude of the wave well.
			}
	\end{minipage}
	\hfill
	\begin{minipage}{.46\textwidth}
		\centering
		\includegraphics[scale=0.43]{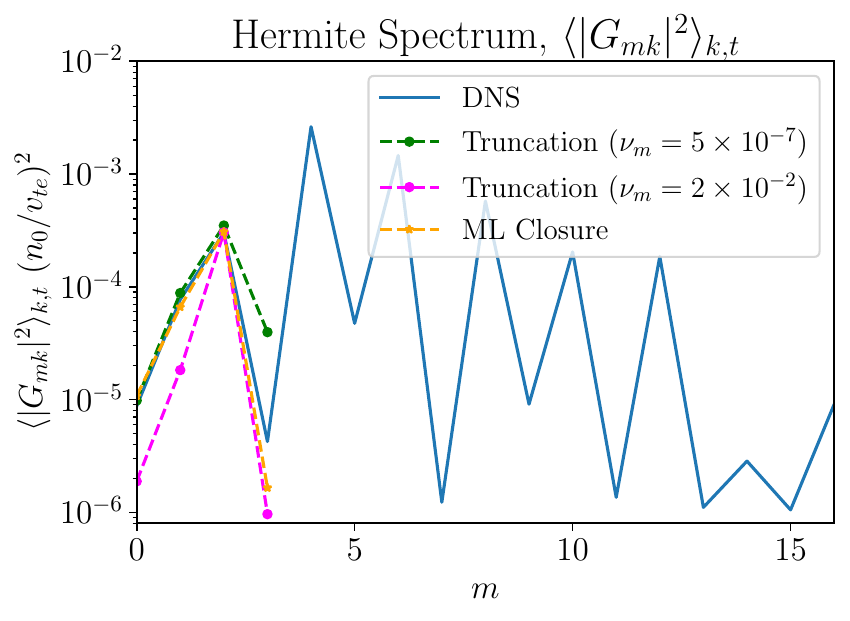}
		\caption{
			\label{fig:NL_Hermite_spectrum}
			Hermite spectra of the high-velocity-resolution ($M=65$), truncated ($M=4$) simulations, and ML closure for the initial value problem in Figure \ref{fig:NL_baseline}\, averaged over $k$ and $t$.  While both the ML closure model and truncated simulation agree with the high-resolution direct numerical simulation, the ML closure model shows closer agreement.}
	\end{minipage}
\end{figure}

Finally, we create an ML closure model for the strongly nonlinear regime of the Vlasov-Poisson system.  
When the amplitude of the initial density perturbation becomes significant relative to the background, the nonlinear term dominates over the energy cascade in linear Landau damping. 
Figure \ref{fig:NL_baseline} presents the numerical solution to the initial value problem from Section \ref{sec:weakly_NL_VP}\,, with $\epsilon$ increased to 0.18.  
When compared to Figure \ref{fig:linear_landau}\,, the behavior at high amplitude deviates from the linear theory.  
Additionally, a convergence study indicated that more Hermite moments ($M=65$) are necessary to accurately resolve the low-order spectrum than in the weakly nonlinear case.
(For more detail on this, see Appendix \ref{sec:hermite_convergence}\,.)
We set $\nu_m$ to $5 \times 10^{-7}$, following the procedure we used to determine the baseline hypercollisional regularization parameter for the linearized and weakly nonlinear cases.
From the start of the simulation until approximately $t=25$, the mode damps at a faster rate than in the linear theory.  
After the initial transient, the mode saturates to a steady state.  
When the ML closure model is introduced, the lower Hermite resolution ($M=4$) system maintains an accurate frequency and slightly overdamps the amplitude.
The training phase for the ML closure model was set to $50 \leq t < 100$ to avoid the most severe initial transient behavior, and the time trace for the ML closure begins at $t=100$.

\begin{figure}
	\centering
	\includegraphics[scale=0.45]{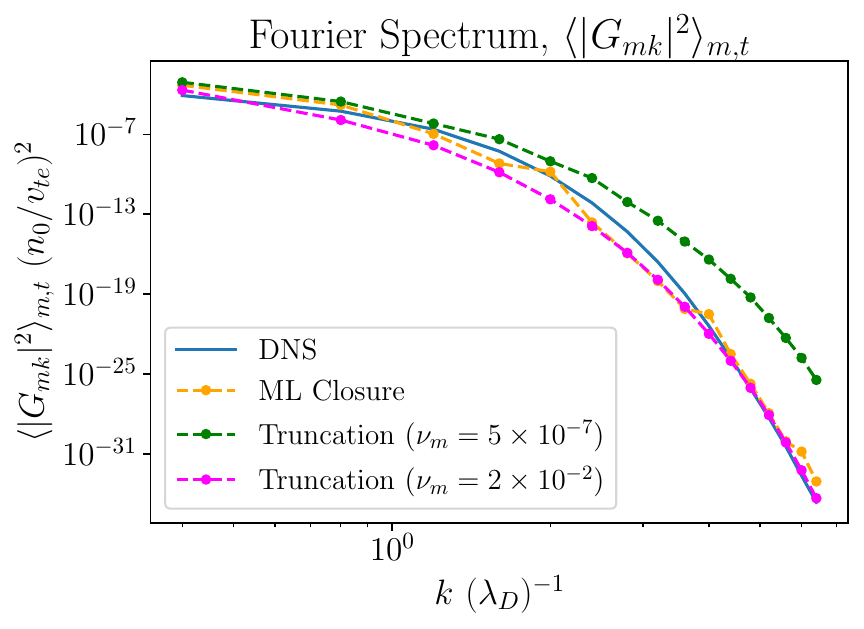}
	\caption{
		\label{fig:NL_Fourier_spectrum}
		Fourier spectra of the simulations in Figure \ref{fig:NL_Hermite_spectrum}\,.  The ML closure resolves the wavenumber spectrum more accurately than the truncated simulations, improving simulation accuracy at low resolution in velocity space.  
	}
\end{figure}

As in the previous section, Figure \ref{fig:NL_Hermite_spectrum} displays a comparison between the Hermite spectra of the high-resolution baseline, the ML closure model, and two truncated low-resolution simulations, where one has an increased amount of hypercollisional regularization to account for the lower resolution.  
At this higher initial amplitude, the Hermite spectrum of the baseline does not decrease monotonically, creating a more challenging scenario for the reservoirs to model. 
At the same value of $\nu_m$, both the truncated low-resolution simulation and the ML closure model accurately capture the Hermite spectrum of the system at low $m$, but as Figure \ref{fig:NL_Fourier_spectrum} reveals, only the ML closure model also agrees with the Fourier spectrum.  
Mean-squared errors in the ML closure spectra are $3.66 \times 10^{-18}$ and $1.19 \times 10^{-14}$ for the Hermite and Fourier spectra, respectively.
The ML closure successfully captures both spectra of the system, reducing the required velocity-space resolution by a factor of sixteen. 
We observe some oscillatory behavior in the Fourier spectrum of the ML closure, but our tests do not identify a mechanism that causes this behavior.
We obtain the reservoir hyperparameters by the same method as the previous section, adjusting them to $\beta = 10^{-6}$, $T=50$, and 120 nodes in each reservoir.  Other parameters remain the same.

A potentially surprising result is that the Hermite spectrum of the na\"{i}vely truncated simulation is more accurate than its Fourier spectrum, despite the fact that the truncation occurs in Hermite space.  
We intuit that this result is due to a pile up of excess free energy at the closure moment, which then nonlinearly advects in Fourier space.
Adjusting the strength of the hypercollisions slightly improves the accuracy of the Fourier spectrum, at the cost of overdamping the density and momentum moments.
An analysis of the flux of free energy in this system similar to the work of \citet{Meyrand19} may be used to investigate this result further, but that is beyond the scope of this work.

\section{Summary and Conclusion}
\label{sec:conclusion}
In this paper, we have used reservoir computing to present an ML, velocity-space closure model for the hierarchy of Hermite moments in the one-dimensional Vlasov-Poisson system.  
The closure model serves as a proof-of-principle that reservoir computing can be used to reduce the simulation domain requirements for studying plasma dynamics.  

A logical objective to pursue is to extend the closure to reduce the required resolution in the Fourier representation of position space.  
In practice, gyrokinetic turbulence simulations commonly use significantly more resolution for configuration space than they do for velocity space, in part because of the near-Maxwellian nature of the velocity distributions.  
Therefore, a Fourier-space closure model would potentially be more desirable than a velocity-space closure.  
However, the structure of the nonlinear term poses a challenge to achieving that goal.  
In the Vlasov-Poisson system, the coupling between resolved and unresolved scales in velocity space occurs as a sum of nearest-neighbors in Hermite space. This locality in the spectral domain defines a clear, single closure term.  
In contrast, the convolution in the nonlinear term establishes nonlocal interactions across scales in Fourier space.  
An ML closure for Fourier space would require a different structure than the one developed for this paper.  

The eventual goal is to build closure models that can reduce the domain requirements for higher dimensional turbulence simulations, including the full gyrokinetic equation.  
A next step toward that goal would be to test this velocity-space closure in that system.
There has been recent interest in applying spectral formulations of velocity space to gyrokinetic simulations, particularly with a Hermite basis used for parallel velocity and a Laguerre polynomial basis for perpendicular velocity \citep{Jorge17,Mandell18,Hoffmann23b,Hoffmann23,Frei23,Frei24}.
A successful implementation of an ML closure for Fourier-Hermite-Laguerre codes, including \texttt{Gyacomo} \citep{Hoffmann23} and \texttt{GX} \citep{Mandell24} may improve their capabilities to resolve turbulence statistics with lower resolution requirements.
Empirically, both codes calculate turbulent heat fluxes with high accuracy using eight or fewer Laguerre modes, suggesting that future work on ML closures should prioritize reductions in spatial and parallel velocity resolution requirements.

An ideal closure model would have a compact, symbolic form from which physical intuition can be derived.  
Significant recent progress toward interpretable, symbolic closures has been achieved using sparse regression techniques \citep{Donaghy23,Cheng23,Ingelsten24}.
One challenge that these algorithms face is the difficulty of finding symbolic closures that are not local in space.
The analytic closure derived by \citet{HammettPerkins} for the linear limit of the system in this paper is local in Fourier space, but it includes a nonlocal Hilbert transform when expressed in configuration space.
Closures for the nonlinear form of this system or other systems of interest may also require information that is nonlocal in configuration space, but local in a spectral representation.
The ML closure developed in this paper and by \citet{Huang25} apply a Fourier representation to the data, providing the neural networks direct access to locality in spatial scales, and the velocity-space closure in this paper is also local in Hermite space.
Though interpeting moment closures that use neural networks is more challenging than closures derived by sparse regression techniques, future work may incorporate feature importance identification methods like those formalized by \citet{Lundberg17}.

\section*{Acknowledgements}
We express our gratitude to R. Gaur, N. R. Mandell, and M. Nastac for helpful discussions about turbulence, M. Almanza, E. P. Alves, D. Carvalho, J. Chou, F. Fiuza, G. Guttormsen, B. Jang, N. F. Loureiro, M. McGrae-Menge, J. Pierce, and A. Velberg, for their perspectives on subgrid modeling, and E. Ott, D. Patel, and A. Wikner for sharing their insights on reservoir computing.  

\section*{Funding}
This work was made possible through the support of the United States Department of Energy under contract numbers DE-FG02-93ER54197 and DE-SC0022039.

\section*{Declaration of interest}
M.L. is a consultant for Type One Energy.

\appendix
\section{Hermite Resolution Convergence}
\label{sec:hermite_convergence}
\begin{figure}
	\begin{minipage}{0.46\textwidth}
	\centering
	\includegraphics[scale=0.45]{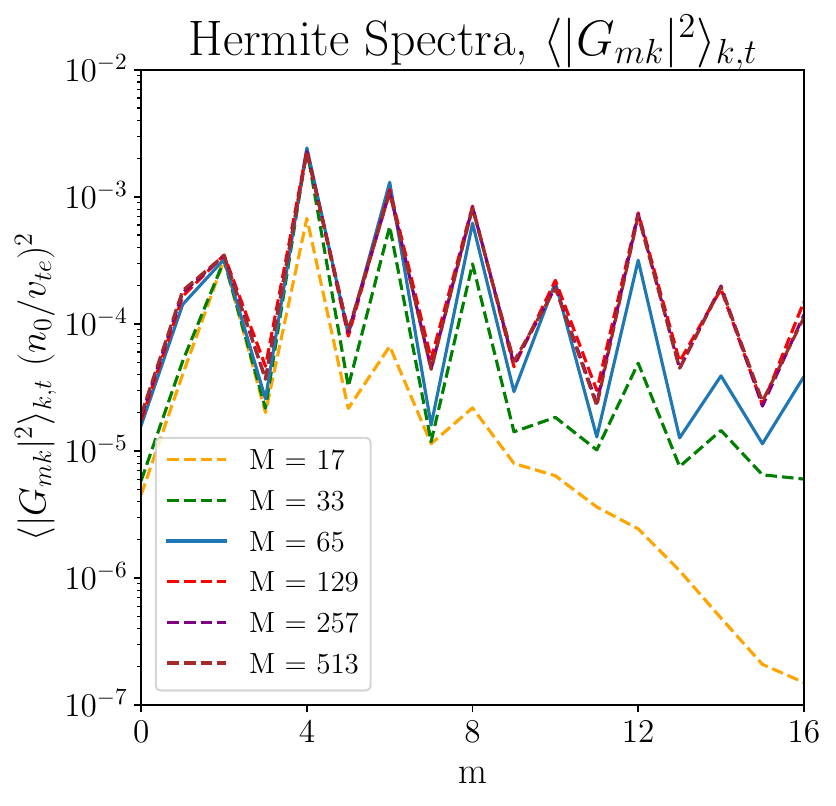}
	\caption{
	\label{fig:NL_Hermite_resolution_scan} 
	Convergence study of the Hermite spectra for the strongly nonlinear initial-value problem from Section \ref{sec:nonlinear_VP}\,. We solve (\ref{eq:full_NL_VP_spectral_start}) - (\ref{eq:full_NL_VP_spectral_end}) without the ML closure at increasing levels of resolution in velocity space.  We selected $M=65$ moments for the high-resolution baseline case in Section \ref{sec:nonlinear_VP}\,, as that case is the first to show convergence in the density and momentum moments.  
	}
	\end{minipage}
	\hfill
	\begin{minipage}{0.46\textwidth}
	\centering
	\includegraphics[scale=0.45]{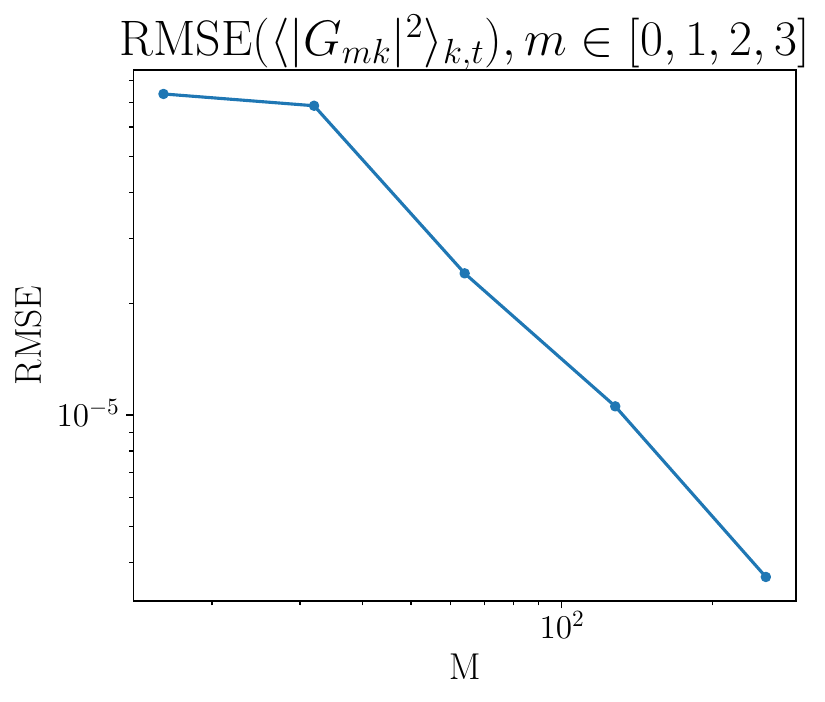}
	\caption{
	\label{fig:low_order_Hermite_rmse}
	Root-mean-square error in the low-order ($m \in [0,1,2,3]$) Hermite spectra at the resolutions plotted in Figure \ref{fig:NL_Hermite_resolution_scan}\,, as compared to the $M=513$ case. The $M=17$ and $M=33$ cases have similar errors, and exponential convergence occurs afterward.}
	\end{minipage}
\end{figure}
In Section \ref{sec:nonlinear_VP}\,, we demonstrated that reservoir computing can be used to construct an ML closure that reduces number of moments required to capture the low-order spectrum of the strongly nonlinear regime of the Vlasov-Poisson system by a factor of sixteen.  In Figure \ref{fig:NL_Hermite_resolution_scan}\,, we present the convergence study that we used to determine the high-resolution ($M=65$) baseline for that case.  
The low-order moments represent important physical quantities,  including density, momentum, and energy.  
Therefore, accurately resolving those moments is a higher priority than resolving the fine-scale structure in velocity-space that the high-order moments capture.  
In Figure \ref{fig:NL_Hermite_resolution_scan}\,, the lowest Hermite resolution that captures the $m=0$ and $m=1$ density and momentum moments within a factor of 2 is $M=65$, so that resolution was selected as the baseline case.  
Figure \ref{fig:low_order_Hermite_rmse} shows a root-mean-square error (RMSE) metric for the low-order Hermite spectra at different Hermite resolutions, where we take the $M=513$ spectrum as ground truth.  Inaccuracy in the density and momentum moments leads to increased RMSE for the $M=17$ and $M=33$ cases, and subsequent errors converge exponentially, beginning with $M=65$.

\section{Time Evolution of Low-Order Moments}
\label{sec:time_traces}

\begin{figure}
	\begin{minipage}{0.46\textwidth}
		\centering
		\includegraphics[scale=0.46]{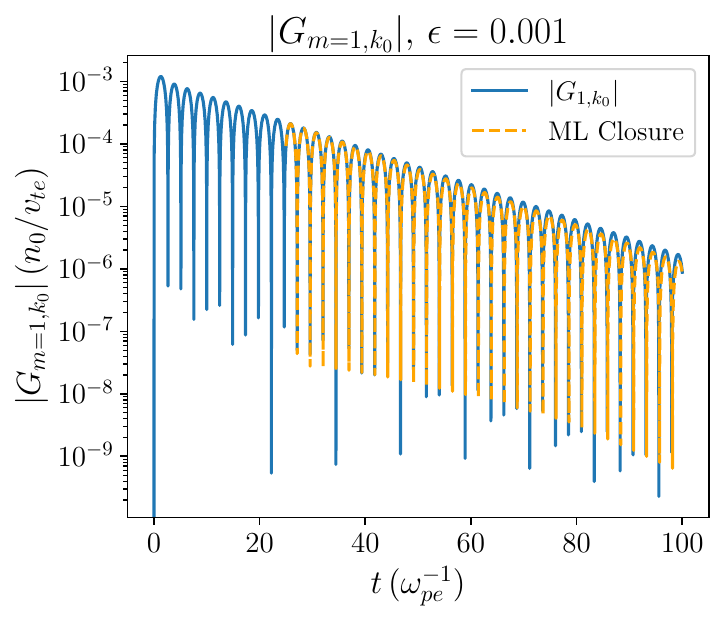}
		\caption{
		\label{fig:weakly_nonlinear_m1}
		Time trace of the $m=1$ moment for the low-amplitude initial-value case in Section \ref{sec:weakly_NL_VP}\,.
		}
	\end{minipage}
	\hfill
		\begin{minipage}{0.46\textwidth}
		\centering
		\includegraphics[scale=0.46]{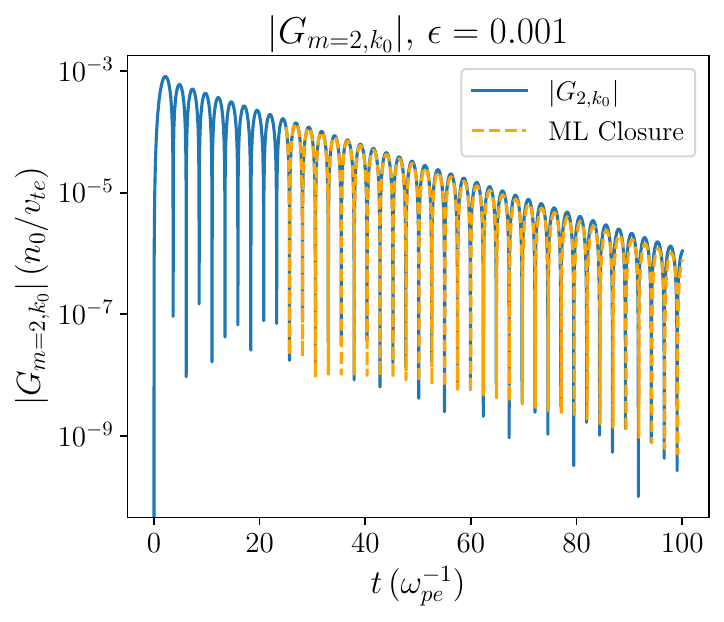}
		\caption{
		\label{fig:weakly_nonlinear_m2}
		Time trace of the $m=2$ moment for the low-amplitude initial-value case in Section \ref{sec:weakly_NL_VP}\,.
		}
	\end{minipage}	
\end{figure}

\begin{figure}
	\begin{minipage}{0.46\textwidth}
		\centering
		\includegraphics[scale=0.46]{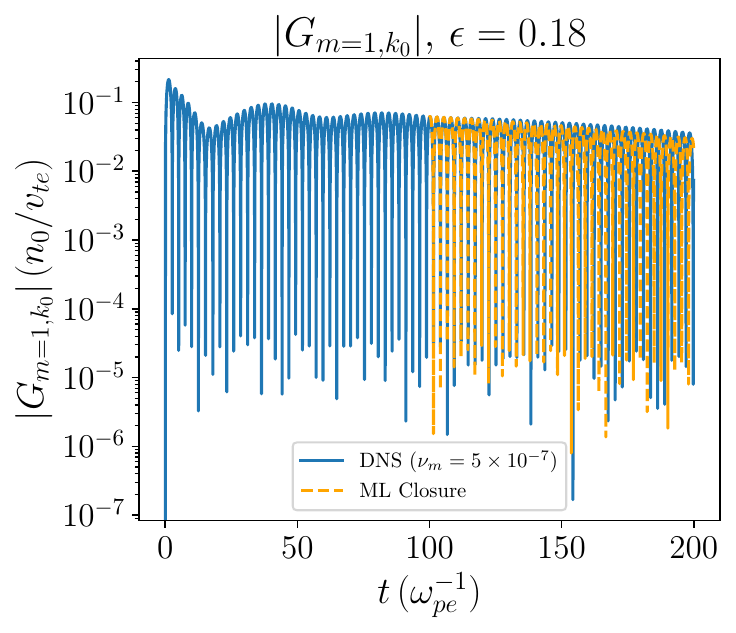}
		\caption{
		\label{fig:nonlinear_m1}
		Time trace of the $m=1$ moment for the high-amplitude initial-value case in Section \ref{sec:nonlinear_VP}\,.
		}
	\end{minipage}
	\hfill
	\begin{minipage}{0.46\textwidth}
		\centering
		\includegraphics[scale=0.46]{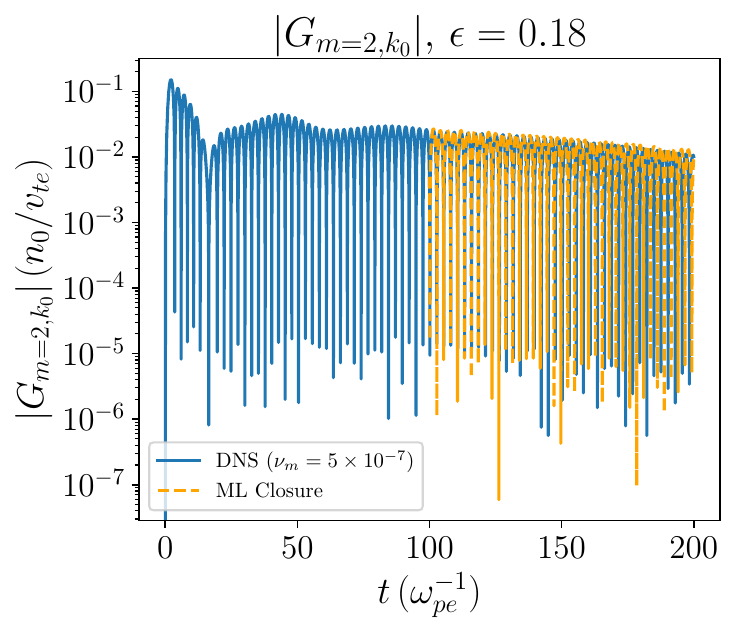}
		\caption{
		\label{fig:nonlinear_m2}
		Time trace of the $m=2$ moment for the high-amplitude initial-value case in Section \ref{sec:nonlinear_VP}\,.
		}
	\end{minipage}	
\end{figure}

Sections \ref{sec:weakly_NL_VP} and \ref{sec:nonlinear_VP} demonstrate that the ML closure can capture spectra of the Vlasov-Poisson system and that the time evolution of the density moment is accurate.
Figures \ref{fig:weakly_nonlinear_m1}$\,-\,$\ref{fig:nonlinear_m2} depict time traces of the $m=1$ and $m=2$ moments for the cases in those sections.
The ML closure resolves the frequencies of the oscillations well. It also successfully captures the amplitude peaks soon after training ends, but it fails to match many of the minima. 
At long times after training, the amplitudes resulting from the ML closure begin to deviate slightly from the DNS amplitudes.
In the weakly nonlinear regime presented in Figures \ref{fig:weakly_nonlinear_m1} and \ref{fig:weakly_nonlinear_m2}\,, the ML closure leads to a slightly lower amplitude at the end of the simulation for both the $m=1$ and $m=2$ moments.
In the strongly nonlinear regime in Figure \ref{fig:nonlinear_m1}\,, the ML closure trends toward a lower amplitude for $m=1$ than in the DNS baseline.  
However, for $m=2$ in that regime, Figure \ref{fig:nonlinear_m2} shows that ML closure trends toward a slightly higher amplitude than the DNS. 
To quantitatively analyze this behavior, we report damping rates for the DNS and ML closure in table \ref{table:damping_rates}\,.
Damping rates were calculated by first extracting the local maxima from each time trace and then solving a linear regression problem for the natural logarithm of those maxima.
An interesting observation is that the $m=1$ and $m=2$ results are identical for the low-amplitude case.
In contrast, in the high-amplitude case, the DNS has a faster damping rate for $m=2$ than for $m=1$.
In the strongly nonlinear regime, the ML closure yields similar damping rates for $m=1$ and $m=2$, but they diverge from the DNS damping rates.

\begin{table}
\begin{center}
{\begin{tabular}{lccc}
		\textit{\textbf{$\epsilon$}} & \textit{\textbf{m}}  & \textbf{DNS damping rate} \textit{\textbf{$(\omega_{pe})$}} & \textbf{ML closure damping rate} \textit{\textbf{$(\omega_{pe})$}} \\[6pt]
		0.001 & 1 & $-6.581 \times 10^{-2}$ & $-6.911 \times 10^{-2}$ \\
		0.001 & 2 & $-6.581 \times 10^{-2}$ & $-6.911 \times 10^{-2}$ \\
		0.18  & 1 & $-5.476 \times 10^{-3}$ & $-7.572 \times 10^{-3}$ \\
		0.18  & 2 & $-8.732 \times 10^{-3}$ & $-7.518 \times 10^{-3}$ \\[6pt]
\end{tabular}}
\caption{
	\label{table:damping_rates}
	Damping rates calculated from the initial-value simulations in Figures \ref{fig:weakly_nonlinear_m1}$\,-\,$\ref{fig:nonlinear_m2}\,.  Here, $\epsilon$ is the amplitude of the initial cosine density perturbation, and $m$ is the Hermite moment number.  Damping rates were calculated by solving a linear regression problem for the natural logarithm of the local maxima in each time series.}
\end{center}
\end{table}

\bibliographystyle{jpp}

\bibliography{closure_refs}

\end{document}